\documentclass[aip, apl, reprint]{revtex4}
\usepackage{amsmath,amssymb,graphicx}
\usepackage{dcolumn}
\usepackage{bm}
\pdfoutput=1

\begin{document}

\preprint{AIP/123-QED}

\title{Co-existence of size-dependent and size-independent thermal conductivities in single layer black phosphorus}
\author{Liyan Zhu}
\affiliation{Department of physics, Center for Computational Science and Engineering, and Graphene Research Center, National University of Singapore, Singapore, 117542, Republic of Singapore}
\author{Gang Zhang}
\affiliation{Institute of High Performance Computing, Singapore 138632, Republic of Singapore}
\author{Baowen Li}
\email{phylibw@nus.edu.sg}
\affiliation{Department of physics, Center for Computational Science and Engineering, and Graphene Research Center, National University of Singapore, Singapore, 117542, Republic of Singapore}
\affiliation{NUS Graduate School for Integrative Sciences and Engineering, National University of Singapore, Singapore, 117546, Republic of Singapore}
\affiliation{Center for Phononics and Thermal Energy Science, School of Physics Sciences and Engineering, Tongji University, Shanghai 200092, Peoples's Repulic of China}

\date{\today}

\begin{abstract}
Thermal conductivity of single layer black phosphorus (BP) is investigated by combining density functional calculations and Peierls-Boltzmann transport equation. Differing from isotropic and divergent thermal conductivities in two-dimensional graphene and MoS$_2$, an compelling co-existence of size-dependent and size-independent thermal conductivities are discovered for single layer BP along zigzag (ZZ) and armchair (AM) direction, respectively. Besides, thermal conductivities in single layer BP are found to be highly anisotropic because of orientation dependent group velocities, e.g., thermal conductivities at 300 K are 83.5 and 24.3 W/m-K along ZZ and AM directions for single layer BP with a size of 10 $\mu m$, respectively. 
\end{abstract}

\pacs{63.20.dk, 63.20.kg, 63.22.Np, 66.70.Df}
\keywords{Black phosphorus, 2D material, Thermal conductivity}
\maketitle

\section{\label{sec:introduction}Introduction}
Graphene, the thinnest two-dimensional (2D) material, has attracted enormous interest since its first exfoliation\cite{Novoselov2004} due to extraordinary mechanical, physical and chemical properties, such as  peculiar band structure, extremely high carrier mobility,  and giant thermal conductivity\cite{Neto2009}. These exceptional properties render graphene a promising material for nanoelectronics. Unfortunately, the lack of band gap limits its application in semiconducting devices. Although there are many ways to open a sizable band gap in graphene\cite{Jariwala2011}, the band gap opening is accompanied by greatly reduced carrier mobility as compared to pristine graphene \cite{Wei2009}. The great success achieved in synthesizing graphene as well as its fantastic and diverse properties have motivated both experimentalists and theoreticians to explore other alternatives of 2D materials\cite{Novoselov2004,Wang2012} to overcome the weakness of graphene.  Among them, MoS$_2$, exhibiting a large band gap, has received much attention. Field effect transistor based on MoS$_2$ demonstrates a giant on-off current ratio, i.e. $ > 10^8$, and near-idea subthreahold swing (50-70 meV/dec)\cite{Kim2012}. However, carrier mobility in single layer MoS$_2$ is much smaller than that in graphene. The initially measured carrier mobility is as low as 0.5-3 $cm^2 V^{-1}s^{-1}$.\cite{Novoselov2005}  By improving fabrication condictions, it could be enhanced to around 200-400 $cm^2 V^{-1}s^{-1}$\cite{Kim2012,Perera2013}. 

Very recently, experimentalists successfully obtained another emergent 2D material, single and few layer black phosphorus (BP)\cite{Li2014,Liu2014}.  BP, the most stable allotrope of phosphorus at ambient condition, is a layered material with a direct band gap of 0.3 eV for bulk\cite{BP_gap}. The band gap, however, gradually increases with the decreasing number of layers in few layer BP films. Several groups have demonstrated the high performance of BP based field effect transistor\cite{Li2014,Liu2014}. For example, Li et al.\cite{Li2014} identified that the 10 nm thick few layer BP exhibits a highest carrier mobility, $\sim$ 1000 $cm^2V^{-1}s^{-1}$. The on-off current ratio in BP based field effect transistor is as high as $10^5$.\cite{Li2014,Liu2014}  Moreover, the electronic transport properties in few layer BP are found to be orientation dependent\cite{Liu2014,Xia2014}. More precisely, carrier mobility, electronic and optical conductivity along armchair (AM) direction are larger than those along zigzag (ZZ) direction\cite{Xia2014}. The sizable and tunable band gap, high carrier mobility, and large on-off current ratio make single and few layer BP ideal in a variety of applications in nanoelectronics and optoelectronics.  However, the heat management has been becoming a critical challenge for electronic devices due to the quick increase of power density. To facilitate heat dissipation, the materials employed in device fabrication must have a good thermal conductivity. Moreover, several groups predicted that monolayer BP might be a potential candidate for thermoelectric devices\cite{lv2014large,Fei2014}. All these potential applications of single layer BP rely on the well understanding on its thermal transport properties, which is still very lacking.

On the other hand, the thermal conductivity demonstrates anomalously length-dependent behaviour for low-dimensional systems, differing from size-independent thermal conductivity predicted by Fourier's law for three-dimensional materials. Theoretical study rigorously proved that the thermal conductivity in 1D Fermi-Pasta-Ulam lattice should follow a power law dependence on the length\cite{LiuSha}. Furthermore, many molecular dynamics simulations revealed that the thermal conductivities of one-dimensional carbon nanotubes\cite{Zhang2005,Zhang2005a,Maruyama2002} and Si nanowires\cite{Yang2010} obey a power law dependence on their length. As for 2D materials, the non-equilibrium MD simulation found a logarithmically divergent thermal conductivity for the 2D nonlinear lattice system\cite{Wang2012a}. Moreover, Lindsay and co-workers\cite{Lindsay2014} theoretically found that the thermal conductivity in graphene is length (L) dependent. Nika et al.\cite{Nika2012} further predicted that the thermal conductivity of graphene should exhibit a logarithmically divergent with respect to the length when L $< 30 \mu m$, which has been experimentally verified by Xu et al\cite{Xu2014}. Furthermore, the theoretical study on single layer MoS$_2$\cite{Li2013} also reveals a length-dependent thermal conductivity similar to the graphene case\cite{Lindsay2014,Nika2012,Xu2014}. So it would be interesting to examine whether the thermal conductivity of single layer BP would diverge with its length or not. In this study, the lattice dynamic properties and thermal conductivity of monolayer BP are investigated by using first principle method and Peierls-Boltzmann transport equation (PBTE). The understanding of thermal transport properties in monolayer BP would also be useful in engineering its thermal conductivity for thermoelectric applications. 

\begin{figure}
\begin{center}
\includegraphics[width=8.5cm]{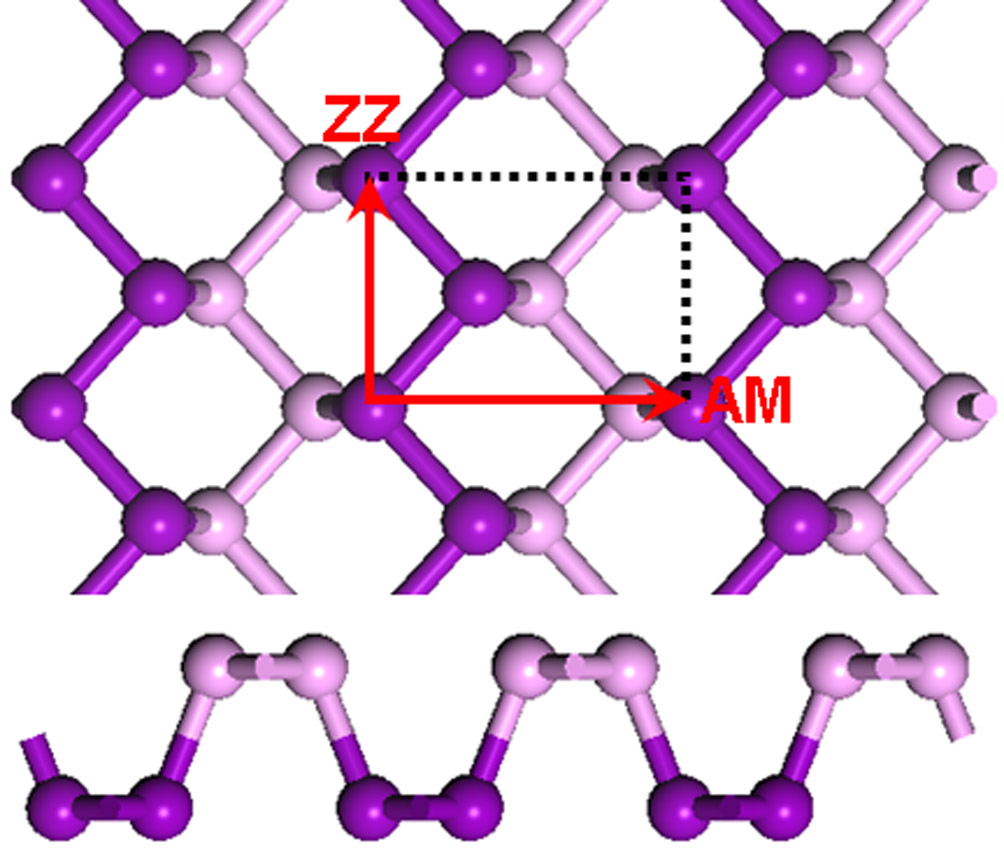}
\caption[struct]{Top and side view of single layer BP. Red arrows represent lattice vectors of primitive cell along ZZ and AM directions. Heavy and light colors denote phosphorus atoms in different plane.}
\label{fig:structure}
\end{center}
\end{figure}

\section{\label{sec:method}Method}
First principle calculations are carried out by using the Vienna Ab initio simulation package\cite{VASP} (VASP). The exchange and correlation interactions between electrons are described by Perdew, Burke, and Ernzerhof (PBE) functional\cite{Perdew1996}. The projector augmented wave pseudopotential\cite{PAW} is adopted to model the interaction between electrons and ions. The energy cutoff is chosen to be 400 eV for the expansion of wavefunction by plane wave basis sets. The structure of single layer BP is fully relaxed with a k-point mesh of 25$\times$25$\times$1.  The vertical distance between BP layers is fixed to be 20 \AA \  , which is large enough to make the suspicious interaction between layers negligible. These combined parameters could make the variance of total energy less than 5 meV/atom. The optimal lattice constants are found to be 3.30 and 4.60 \AA\ along ZZ and AM direction, respectively (see Fig. \ref{fig:structure}). These values are in good agreement with previous theoretical predictions \cite{Qiao2014}. 

Thermal conductivity of single layer BP is calculated by using PBTE with relaxation time approximation as implemented in ShengBTE\cite{ShengBTE,Li2012a}, in which thermal conductivity is given by
\begin{equation}
\kappa_{\alpha} = \frac{1}{N_q V} \sum_{\mathbf{q}, j} C_{\mathbf{q}, j} v_{\mathbf{q}, j, \alpha}^2 \tau_{\mathbf{q}, j}, 
\end{equation}
where $N_q$, and V are total number of q-points sampling Brillouin zone and the volume of primitive cell of single layer BP. In computing V, the thickness of single layer BP is chosen to be half of experimental lattice constants for bulk BP\cite{Akai1989}, namely, 5.239 \AA. $C_{\mathbf{q}, j}$,  $v_{\mathbf{q}, j, \alpha}$ and $\tau_{\mathbf{q}, j}$ are specific heat, group velocity along transport direction $\alpha$, and relaxation time of the phonon mode with wavevector $\mathbf{q}$ and polarization $j$. The expression of relaxation time ($\tau_{\mathbf{q}, j}$) can be found in Ref. \cite{ShengBTE}. The calculations of $C_{\mathbf{q}, j}$,  $v_{\mathbf{q}, j, \alpha}$ and $\tau_{\mathbf{q}, j}$ require second and third order force constants (FCs) as inputs. Both second and third order FCs are extracted from density functional theory computations by using finite displacement method\cite{Li2012, ShengBTE}. The second (third) order FCs are determined from a 5$\times$5$\times$1 (4$\times$3$\times$1) supercells with a 2$\times$2$\times$1 (4$\times$4$\times$1) Monkhorst-Pack k-point mesh, respectively. The interaction range of third order FCs is truncated up to 4.4 \AA.  

\section{\label{sec:res}Results and discussion}
\begin{figure}[!htbp]
\begin{center}
\includegraphics[width=8.5cm]{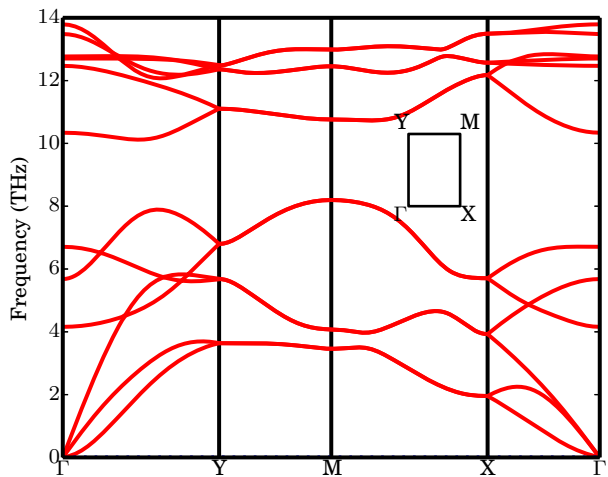}
\caption[dispersion]{Phonon dispersion of single layer BP.}
\label{fig:band}
\end{center}
\end{figure}

Figure \ref{fig:band} shows the phonon dispersion of single layer BP along high symmetrical path of Brillouin zone. The phonon dispersion calculated from first principle is consistent with that obtain from empirical interatomic potential\cite{Kaneta1986}. Similar to graphene, the frequency of out-of-plane acoustic phonon (ZA) exhibits a quadratic dependence on the wavevector as approaching Brillouin zone center, which is a characteristic property of layered materials\cite{Zabel2001}. Based on macroscopic elastic theory of thin plates, the frequency of ZA mode is of the following form\cite{Politano2012}  
\begin{equation}
\omega^2(\mathbf{q}) = A |\mathbf{q}|^4 + B|\mathbf{q}|^2.
\end{equation}
The fitting parameter A is given by $t/\rho_{2D}$, where $t$ and $\rho_{2D}$ are bending rigidity and 2D mass density of single layer BP. According to this equation, the bending rigidity along AM direction is around 1.9 eV, which is slightly larger than that of graphene\cite{Graphene_Bending}; while that along ZZ is estimated to be 6.7 eV. The large bending rigidity along ZZ direction results from its corrugated structure extending along ZZ direction. The group velocities of LA modes along ZZ and AM direction are 8.6 and 4.5 km/s, respectively, which are close to experimental values for bulk BP\cite{Fujii1982}, i.e. 9.6 and 4.6 km/s along ZZ and AM direction respectively.

\begin{figure}[htbp]
\begin{center}
\includegraphics[width=8.5cm]{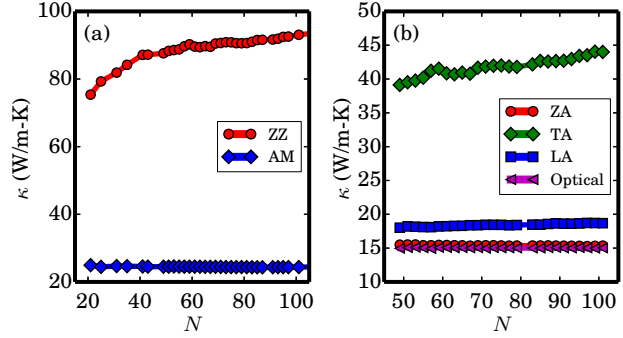}
\caption[kappa]{ Thermal conductivity ($\kappa$) as a function of (a) the size of q-point mesh (N$\times$N$\times$1) and (b) Thermal conductivity along ZZ direction contributed from ZA, TA, LA, and optical phonon branches as a function of N.}
\label{fig:kappa}
\end{center}
\end{figure}

As mentioned in the section of introduction, the 2D systems, e.g., nonlinear 2D lattice\cite{Wang2012a}, graphene\cite{Nika2012,Xu2014} and and MoS$_2$\cite{Li2013}, usually show a length dependent thermal conductivity, which is due to the relaxation times of acoustic phonon modes quickly blow up as approaching the $\Gamma$ point of Brillouin zone\cite{Bonini2012}. The calculated thermal conductivities along both ZZ and AM directions at 300 K are plotted in Fig. \ref{fig:kappa}a. Along AM direction, the thermal conductivity,  $\kappa^{AM}$, is almost a constant (24.3 W/m-K) with respect to the q-mesh size increases. However, the thermal conductivity along ZZ direction ($\kappa^{ZZ}$) steadily increases with the increasing number of q-points. To further understand the dependence of $\kappa^{ZZ}$ on the size of q-mesh, we plot the contribution of ZA, TA, LA, and optical phonon modes to the thermal conductivity along ZZ direction in Fig. \ref{fig:kappa}b. The thermal conductivities contributed by ZA, LA, and all optical phonons do not change with the increasing size of q-mesh. The divergent $\kappa^{ZZ}$ is mainly due to the contribution from TA phonons ($\kappa_{TA}$).

\begin{figure*}[!p]
\begin{center}
\includegraphics[width=17cm]{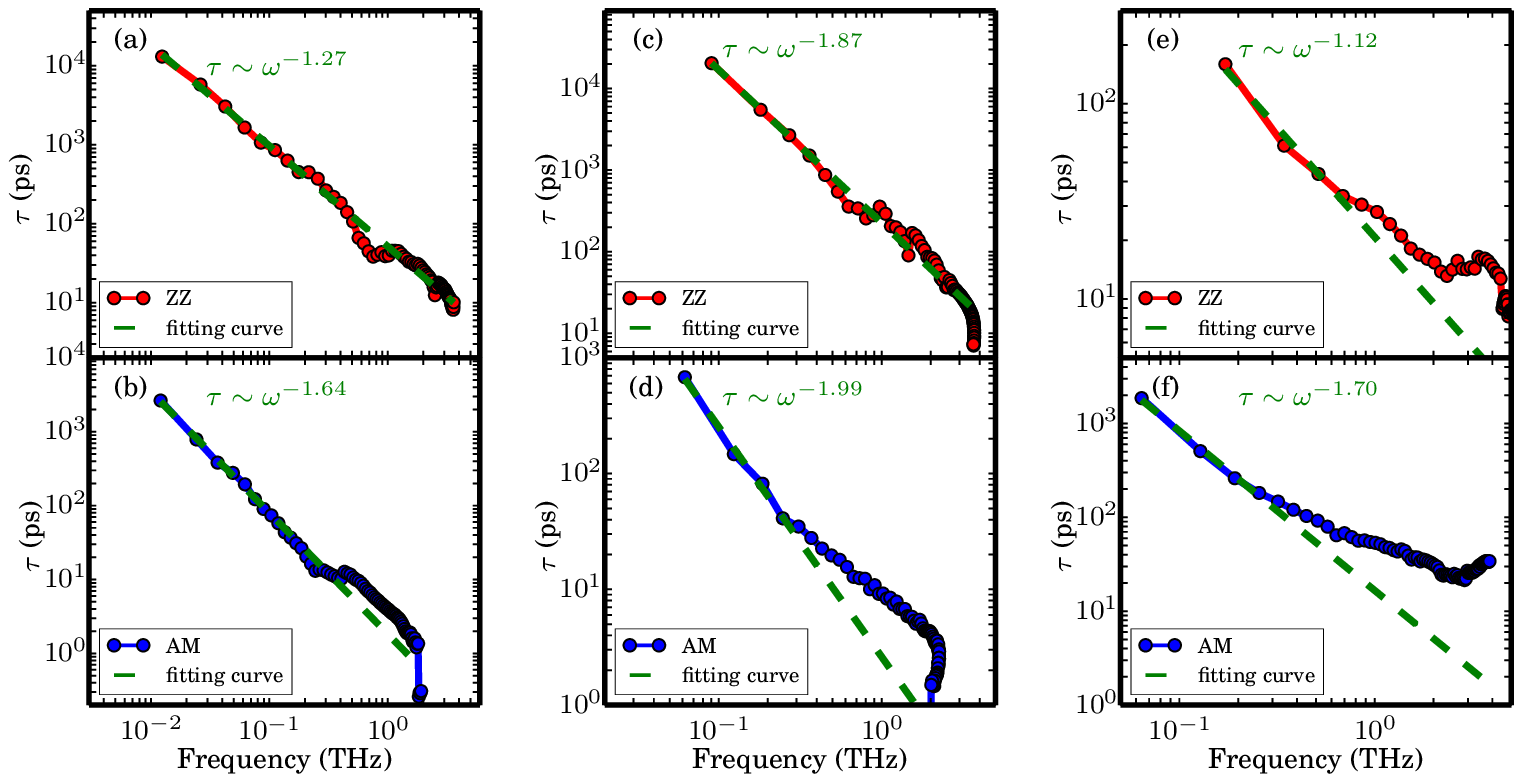}
\caption[ang]{Dependence of phonon relaxation time on frequency for (a-b) ZA, (c-d) TA, and (e-f) LA phonon. Top and bottom panels denote the wavevector $\mathbf{q}$ approaching Brillouin zone center along ZZ ($\Gamma-Y$) and AM ($\Gamma-X$) directions, respectively.}
\label{fig:lifetime}
\end{center}
\end{figure*}

\begin{figure*}[p]
\begin{center}
\includegraphics[width=17cm]{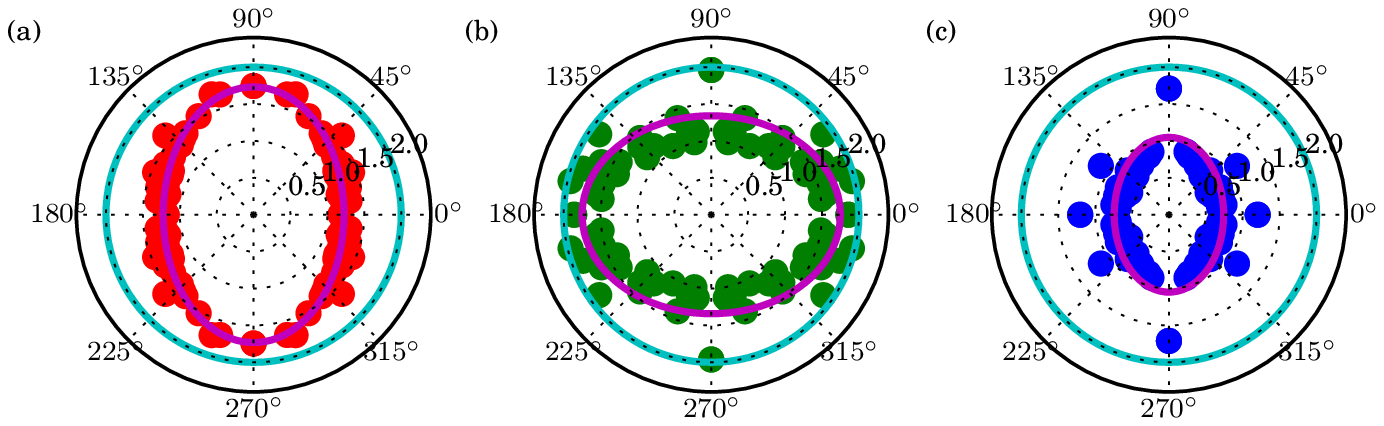}
\caption[ang]{Scaling exponent of phonon relaxation time ($\tau$) on frequency ($\omega$), $n$, as a function of the orientation of wavevector $\mathbf{q}$ approaching Brillouin zone center, namely, $\tau \sim \omega^{-n}$. Exponent $n$ for (a) ZA, (b) TA, and (c) LA phonons are plotted in left, central, and right panels, respectively. The angles of 0 and 90$^\circ$ correspond to ZZ ($\Gamma-Y$) and AM ($\Gamma-X$) directions in Brillouin zone, respectively. The solid lines in cyan denote n=2; while magenta solid lines are plotted to guide the eye. }
\label{fig:lifetime_ang}
\end{center}
\end{figure*}

For 2D material, the thermal conductivity contributed from the vicinity of Brillouin zone center, assuming an isotropic phonon dispersion, can be simply approximated as
\begin{equation}
\label{eq:kappa_integral}
\kappa \sim \sum_{j} \int_{q}\int_{\theta} C (vcos\theta)^2 \tau q dqd\theta  \sim \sum_{j} \int_{q} C v^2 \tau q dq . 
\end{equation}
Since the frequencies of ZA modes show a quadratic dependence on wavevector q, namely, $\omega_{ZA} \sim q^2$. So their group velocities $v_{ZA} \sim q$. However, the frequencies linearly scale with wavevector for both TA and LA modes, i.e. $\omega_{TA/LA} \sim q$. As a consequence, the group velocities of low-frequency TA/LA modes would be constants, namely, $v_{TA/LA} \sim q^0$. While the relaxation times of low-frequency acoustic phonon modes generally follow a power law dependence on frequency(see Fig. \ref{fig:lifetime}), namely, $\tau \sim \omega^{-n}$. Moreover, $C$ approaches to a constant as $\omega \to 0$. Therefore, the integrand in Eq. \ref{eq:kappa_integral} for ZA modes should scales as 
\begin{equation}
\kappa_{ZA} \sim  \int_{q}q^0 q^2 \cdot \omega_{ZA}^{-n} \cdot q dq \sim \int_{q}q^{3-2n} dq 
\end{equation}
 But for TA and LA phonon modes, the integrand would scales as 
\begin{equation}
\kappa_{TA/LA} \sim \int_{q}q^0 q^0 \cdot \omega_{TA/LA}^{-n} \cdot q dq \sim \int_{q}q^{1-n} dq 
\end{equation}
Therefore, the integrals for all three acoustic branches will diverge if and only if $n \geq 2$. However, the exponent $n$ is orientation dependent due to the relatively low symmetry of single layer BP. We plot the exponent $n$ as a function of the orientation of wavevector $\mathbf{q}$ approaching Brillouin zone center in Fig. \ref{fig:lifetime_ang}, in which the angle of 0 means ZZ ($\Gamma-Y$) direction and $90^\circ$ AM ($\Gamma-X$) direction in reciprocal space. It is evident that $n$ for ZA and LA phonon modes are all smaller than 2. Therefore, the thermal conductivity contributed from ZA and LA phonon branches should converge with respect the number of q-points as shown in Fig. \ref{fig:kappa}b. But the exponent for TA phonon modes are very close or equal to 2, especially for the orientation of wavevector $\mathbf{q}$ distributed around ZZ direction. This is the reason why the $\kappa_{TA}$ along ZZ direction will diverge with the size of q-mesh increases. But the group velocities along AM direction are close to 0 for those TA phonon modes. So even though the relaxation times of those phonon modes quadratically depend on their frequencies, their contribution to the $\kappa^{AM}$ is limited. More importantly, most phonon modes with an orientation of wavevector close to AM direction are smaller than 2 except for the wavevector exactly along AM direction. Thus, the $\kappa_{TA}$ along AM direction does not diverge with the increasing number of q-points sampling Brillouin zone. Intuitively, the corrugated structure periodically stacking along AM direction will inhibit the phonon transport along AM direction.

\begin{figure}[htbp]
\begin{center}
\includegraphics[width=8.5cm]{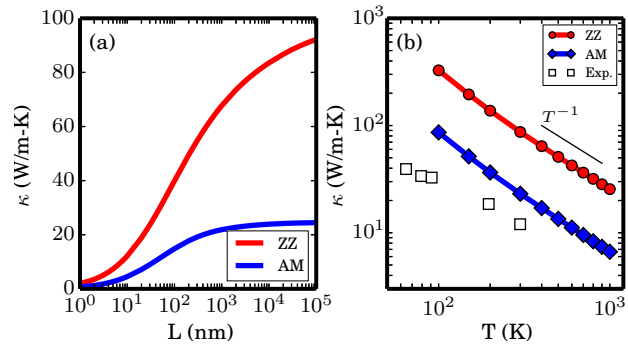}
\caption[kappa_L]{Thermal conductivity ($\kappa$) as a function of length (a) and temperature (b). Open squares in panel b are experimental thermal conductivities for bulk BP\cite{Slack1965}.}
\label{fig:kappa_vs_LT}
\end{center}
\end{figure}

Considering the fact that the BP employed in experimental measurements and device fabrications are of finite size, e.g. a few micrometers\cite{Li2014,Liu2014}, it is important to take phonon-boundary scattering into account when estimating thermal conductivity of finite BP flake. The phonon boundary scattering rate is empirically estimated by 
\begin{equation}
\tau^{-1}_{B}(\mathbf{q}, j) = \frac{2|v_{\alpha}(\mathbf{q}, j)|}{L},
\end{equation}
where L represents the size of sample; $v_{\alpha}$ is group velocity component along heat transport direction. Then, the phonon relaxation time ($\tau$) is calculated as $\tau^{-1} = \tau^{-1}_{Anh} + \tau^{-1}_{B}$, where $\tau_{Anh}$ is phonon relaxation time due to anharmonic phonon-phonon interaction. We plot the thermal conductivities as a function of size in Figure \ref{fig:kappa_vs_LT}a. As clearly seen from Figure \ref{fig:kappa_vs_LT}a, the size strongly affects the thermal conductivity. For example, at a size of 50 nm, the thermal conductivities decrease by more than 50\% for AM directions. So it might be an efficient way to lower thermal conductivity by patterning BP into a nanoribbon to incorporate phonon-boundary scattering. If we further roughen the edges of BP nanoribbon, we may achieve much low thermal conductivities which would facilitate its application in thermoelectric devices.  On the other hand, the thermal conductivity along ZZ direction is about 70 W/m-K at a length of several micrometer which is a typical size of field effect transistor\cite{Li2014,Liu2014}. This value is slightly larger than that of bulk germanium (about 60 W/m-K\cite{kappa_Ge}), which implies that the thermal transport property of single layer BP is not a constraint for its application as semiconductor devices because the accumulated heat can be efficiently dissipated along ZZ direction.

Figure \ref{fig:kappa_vs_LT}b illustrates thermal conductivity as a function of temperature in the range of 100 to 1000 K for single layer BP with a size of 10 $\mu m$. The thermal conductivities drop steadily with the increasing temperature and obey a power law dependence on temperature with an exponent of -1 for both ZZ and AM directions at high temperature, namely,  $\kappa \sim T^{-1}$. This temperature dependence is consistent with Eucken's law\cite{Eucken1911,Eucken1913}, which is due to that the scattering rate of dominant Umklapp process at high temperature is proportional to temperature\cite{Holland1963}. The thermal conductivities from our theoretical calculations show a similar trend as experimental measurements on bulk BP\cite{Slack1965}. But our theoretical values are larger than bulk ones, which might due to two factors. Firstly, the interlayer van deer Waals interaction, absent in single layer BP, might reduce the thermal conductivity of bulk BP. For example, the thermal conductivity in few layer graphene quickly drops with increasing number of layers. Secondly, the experimentally measured sample may contain some extrinsic defects, e.g. vacancy defects, substitutional defects, and intercalated atoms. These defects would greatly reduce the thermal conductivity.

\begin{figure}[tbp]
\begin{center}
\includegraphics[width=8.5cm]{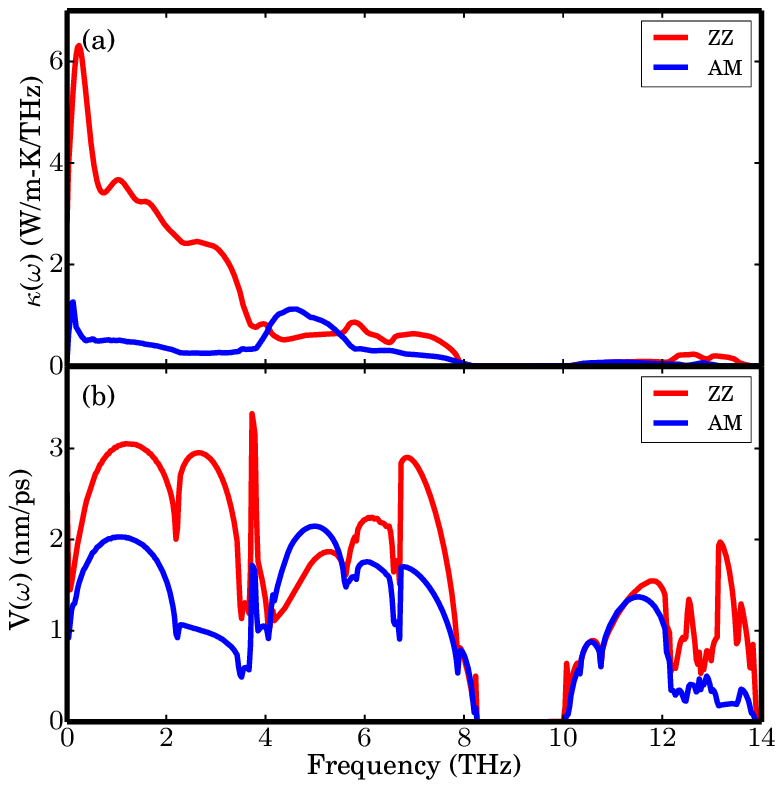}
\caption[omega solved kappa]{Omega resolved (a) thermal conductivities ($\kappa(\omega)$) and (b) average group velocities ($V(\omega)$) along ZZ and AM directions.}
\label{fig:kappa_omega}
\end{center}
\end{figure}

Another important feature shown in Figure \ref{fig:kappa}a is the anisotropy in thermal conductivities. Specifically speaking, the thermal conductivity along ZZ direction is much larger than that along AM direction.  For example, the former along ZZ direction is about three times larger than the latter at 300 K. To understand the orientation dependent thermal conductivity in single layer BP, we calculate frequency resolved thermal conductivities ($\kappa(\omega)$) and plot them in Figure \ref{fig:kappa_omega}a. It is evident that the major difference between thermal conductivities along ZZ and AM directions comes from the low-frequency part ($< 4$ THz). This is because the group velocities along AM direction are much smaller than those along ZZ direction within this frequency range (see Fig. \ref{fig:kappa_omega}b). More interestingly, experimental measurements\cite{Xia2014} and theoretical study\cite{Fei2014} both found that the electronic conductivity along AM direction is higher than that along ZZ direction, which is exactly opposite to the trend observed for thermal conductivity. This might be an advantage for the thermoelectric applications. Combining the large electronic conductivity and low thermal conductivity along AM direction, the BP nanoribbons patterned along AM direction might reach relatively high thermoelectric efficiency. However, the intrinsic thermal conductivity along AM direction of monolayer BP is still larger than expected value to achieve high ZT for thermoelectric applications\cite{lv2014large}. Some other techniques, e.g. strain, defect, and confinement effect, might be employed to suppress the thermal conductivity further. 

\section{\label{sec:conclusion}Conclusion}

In conclusion, we have carefully studied the thermal transport properties of single layer BP by first principle method and PBTE. An intriguing co-existence of size-dependent and size-independent intrinsic thermal conductivities is obtained for phonons transporting along ZZ and AM direction in single layer BP, respectively, which is quite different from isotropically divergent thermal conductivity observed in 2D graphene\cite{Lindsay2014,Nika2012,Xu2014} and MoS$_2$\cite{Li2013}. This is mainly due to the phonon relaxation time of low-frequency TA phonon modes almost quadratically depends on frequency for those phonon modes with an orientation of wavevector close to ZZ direction. Besides, the thermal conductivities in single layer BP is highly anisotropic. The $\kappa^{ZZ}$ is almost three times larger than that $\kappa^{AM}$ at 300 K. The anisotropy in thermal conductivity originates from orientation dependent group velocities. 

\section{Acknowledgements} 
This work is supported in part by the Ministry of Education (MOE), Singapore, by Grant MOE2012-T2-1-114. We thank the  computing resources at the A*STAR Computational Resource Centre, Singapore.

\bibliographystyle{aipnum4}
%

\end{document}